\documentstyle[11pt,moriond,epsfig]{article}

\def\be{\begin{equation}}
\def\ee{\end{equation}}
\def\bea{\begin{eqnarray}}
\def\eea{\end{eqnarray}}

\begin{document}
\title{QUANTUM COMPUTING AND QUANTUM COMMUNICATION\\
WITH ELECTRONS IN NANOSTRUCTURES}

\author{Daniel Loss, Guido Burkard and Eugene V. Sukhorukov}

\address{Department of Physics and Astronomy, University of Basel,\\
Klingelbergstrasse 82, CH-4056 Basel, Switzerland}

\maketitle\abstracts{
If the states of spins in solids can be created, manipulated, and
measured at the single-quantum level, an entirely new form of
information processing, quantum computing and quantum communication, will be
possible.
We review our proposed spin-quantum dot architecture for a
quantum computer and review some recent results on a deterministic
source of entanglement generated by coupling quantum dots.
Addressing the feasibility of quantum communication with entangled
electrons we
consider a scattering set-up with an entangler and beam splitter where the
current
noise exhibits bunching behavior for electronic singlet states and
antibunching
behavior for triplet states.
We  show that spin currents can produce noise even
in the absence of any charge currents.
}

\section{Introduction}

Quantum computation (QC) has attracted much
interest recently \cite{Steane98}, as it opens up the possibility
of outperforming  classical computation through new and more powerful
quantum algorithms such as the ones discovered by Shor \cite{Shor94} and
by Grover \cite{Grover}. There is now a growing list of quantum tasks
\cite{DiVincenzoLossMMM} such as
secret sharing, error correction schemes,  quantum
teleportation, {\it etc.},  that have indicated even more the desirability
of experimental implementations of QC.
In QC the state of
each bit is allowed to be any state of a quantum two-level
system---a qubit, and QC proceeds by one and two-qubit
gates by which all quantum algorithms can be implemented \cite{Barenco}.
There are now a
number of  schemes which have been proposed to realize physical
implementations of qubits and quantum gates
\cite{Cirac95,Cory97,Gershenfeld97,Loss98,Schon,Averin,Ioffe,Kane}.
Here we will review
our qubit proposal \cite{Loss98} based on the spin of electrons confined in
quantum dots (or other confined structures such as molecules or atoms). One
of
the essential features of such qubits is that they are scalable to many
qubits, that recent experiments demonstrated very long spin  decoherence
times
in semiconductors \cite{Kikkawa97},
and that the qubit defined as electron-spin is mobile and thus can be used
for
implementing quantum communication schemes \cite{DiVincenzoLossMMM}.

We  first review our recent results on quantum gates and quantum dots
by which entanglement can be generated \cite{Loss98,Burkard}, and then we
describe a set-up by which
such entanglement can be observed in transport/noise
measurements \cite{NoiseEntangle}.

\section{Quantum Computing: Electron Spin as Qubit in Quantum Dots}

Semiconductor quantum dots, sometimes referred to as artificial atoms,
are small devices in which charge carriers are confined in all three
dimensions \cite{jacak}. The confinement is usually achieved by
electrical gating and/or etching techniques applied e.g. to a
two-dimensional electron gas (2DEG).
In GaAs heterostructures the number of electrons in the quantum dots can be
changed one by one starting from zero \cite{tarucha}. Typical
magnetic fields ($B\approx 1\,{\rm T}$) correspond to
magnetic lengths on the order of $l_B\approx 10\,{\rm nm}$, being of the
same size
as quantum dots. As a consequence, the dot spectrum depends strongly on the
applied magnetic field \cite{jacak,kouwenhoven,ashoori}.
In coupled quantum dots which can be considered  as
artificial molecules, Coulomb blockade effects \cite{waugh} and
magnetization \cite{oosterkamp} have been observed as well as the
formation of a delocalized ``molecular state'' \cite{blick}.

In addition to a well defined qubit such as the electron spin considered
here, we also
need a deterministic ``source of entanglement", i.e.  a mechanism by which
two
specified qubits at a time can be entangled  so as to produce
the fundamental quantum XOR (or controlled-NOT) gate operation,
represented by a unitary operator $U_{\rm XOR}$ \cite{Barenco}.  This can be
achieved by temporarily coupling two spins \cite{Loss98}.  In short, due to
the
Coulomb interaction and the Pauli exclusion principle the ground state of
two
coupled electrons is a spin singlet, i.e. a highly entangled spin state.
This
physical picture translates into an exchange coupling $J(t)$ between the two
spins ${\bf S}_{1}$ and ${\bf S}_{2}$ described by a Heisenberg
Hamiltonian
\begin{equation}\label{Heisenberg}
H_{\rm s}(t)=J(t)\,\,{\bf S}_1\cdot{\bf S}_2.
\end{equation}
If the exchange coupling is pulsed such that $\int dtJ(t)/\hbar =
J_0\tau_s/\hbar = \pi$ (mod $2\pi$), the associated unitary time
evolution $U(t) = T\exp(i\int_0^t H_{\rm s}(\tau)d\tau/\hbar)$
corresponds to the ``swap'' operator $U_{\rm sw}$ which simply
exchanges the quantum states of qubit 1 and 2 \cite{Loss98}. Furthermore,
the quantum XOR can be obtained \cite{Loss98} by applying the sequence
$\exp(i(\pi/2)S_1^z)\exp(-i(\pi/2)S_2^z)U_{\rm sw}^{1/2}\exp(i\pi
S_1^z)U_{\rm sw}^{1/2}\equiv U_{\rm XOR}$, i.e. a combination of
``square-root of swap'' $U_{\rm sw}^{1/2}$ and single-qubit rotations
$\exp(i\pi S_1^z)$, etc. Since $U_{\rm XOR}$ (combined with
single-qubit rotations) is proven to be a universal quantum
gate \cite{Barenco}, it can therefore be used to
assemble any quantum algorithm.  Thus, the study of a quantum XOR gate
is essentially reduced to the study of the {\it exchange mechanism}
and how the exchange coupling $J(t)$ can be controlled experimentally.
We wish to emphasize that the switchable coupling mechanism described
in the following need not be confined to quantum dots: the same
principle can be applied to other systems, e.g. coupled atoms in a
Bravais lattice, supramolecular structures, or
overlapping shallow donors in semiconductors.
As to the latter example, it has been demonstrated in Ref. \cite{Kane}
that the coupling mechanism proposed by us can also be used to couple
nuclear spins (playing the role of qubits) via a tunable exchange coupling
of the
s-electrons (of P embedded in Si) to which the nuclear spins are coupled via
hyperfine interaction. The main reason to
concentrate here on quantum dots is that these systems are at the center of
many
ongoing experimental investigations in mesoscopic physics, and thus
there seems to be reasonable hope that these systems can be made into
quantum gates functioning along the lines described here.

\subsection{Model for the Quantum Gate with Coupled Quantum Dots}

We consider a system of two laterally coupled quantum dots containing
one (conduction band) electron each \cite{Burkard}. It is
essential that the electrons are allowed to tunnel between the dots,
and that the total wave function of the coupled system must be
antisymmetric. It is this fact which introduces correlations
between the spins via the charge (orbital) degrees of freedom.
The Hamiltonian for the
coupled system is,
$H = \sum_{i=1,2} h_i+C+H_{\rm Z} = H_{\rm orb} + H_{\rm Z}$, where
\begin{eqnarray}
h_i = \frac{1}{2m}\left({\bf p}_i-\frac{e}{c}{\bf A}({\bf r}_i)
\right)^2+V({\bf r}_i),\,\,\,\,\,
C={{e^2}\over{\kappa\left| {\bf r}_1-{\bf r}_2\right|}}\,\,\,\,.
\label{hamiltonian}
\end{eqnarray}
Here, $h_i$ describes the single-electron dynamics
confined to the $xy$-plane, with $m=0.067\, m_e$ being the effective mass
in GaAs and ${\bf S}_i$ the electron spin.
The dielectric constant in GaAs is $\kappa = 13.1$. We allow for
a  magnetic field ${\bf B}= (0,0,B)$ applied along the $z$-axis
and which couples to the electron charge via the
vector potential ${\bf A}({\bf r}) = \frac{B}{2}(-y,x,0)$, and to the spin
via a Zeeman coupling term $H_{\rm Z}$.
The coupling of the dots (which includes tunneling) is modeled by a quartic
potential,
$V(x,y)=\frac{m\omega_0^2}{2}\left(\frac{1}{4 a^2}\left(x^2-a^2
\right)^2+y^2\right)$,
which separates  into two harmonic wells  of
frequency $\omega_0$, one for each dot, in the limit
$2a\gg 2 a_{\rm B}$, where $a$ is
half the distance between the centers of the dots, and
$a_{\rm B}=\sqrt{\hbar/m\omega_0}$
is the effective Bohr radius of a dot.
This
choice for the potential is motivated by the experimental
fact \cite{tarucha} that
the spectrum of single dots in GaAs is well described by a parabolic
confinement potential,
e.g. with $\hbar\omega_0 =3\,{\rm meV}$ \cite{tarucha},  $a_{\rm
B}= 20\,{\rm nm}$.

The (bare) Coulomb interaction between the two electrons is
described by $C$. The screening length $\lambda$ in almost depleted regions
like few-electron quantum dots can be expected to be much larger than the
bulk 2DEG screening length (which is about $40\,{\rm nm}$ in GaAs).
Therefore, $\lambda$ is large compared to the size of the coupled system,
$\lambda\gg 2a\approx 40\,{\rm nm}$
for small dots, and we will consider the limit of unscreened
Coulomb interaction
($\lambda/a\gg 1$).

\subsection{Exchange Interaction in the Heitler-London approach}

At low-temperatures where $kT\ll
\hbar\omega_0$ we can restrict ourselves to the two lowest
orbital eigenstates of $H_{\rm orb}$, one of which is symmetric
(spin singlet) and the other one antisymmetric (spin triplet).
In this reduced (four-dimensional) Hilbert space, $H_{\rm orb}$
can be replaced by the effective Heisenberg spin Hamiltonian
Eq.~(\ref{Heisenberg})
where the exchange energy $J=\epsilon_{\rm t}-\epsilon_{\rm s}$
is the difference between the triplet and
singlet energy. The above model cannot be
solved in an analytically closed form. However, the analogy between atoms
and quantum dots (artificial atoms) provides us with a powerful set of
variational methods from molecular physics for finding $\epsilon_{\rm t}$
and $\epsilon_{\rm s}$.
In the Heitler-London
approximation and making use of the Darwin-Fock solution for the
isolated dots we find
\cite{Burkard},
\begin{equation}\label{J}
J=\frac{\hbar\omega_0}{\sinh\left(2d^2(2b-1/b)\right)}\Bigg[c\sqrt{b}
\Bigg(e^{-bd^2}{\rm I_0}(bd^2)
- e^{d^2 (b-1/b)}{\rm I_0}(d^2
(b-1/b))\Bigg)+\frac{3}{4b}\left(1+bd^2\right)\Bigg],
\end{equation}
where we introduce the dimensionless distance $d=a/a_{\rm B}$, and ${\rm
I_0}$ is the zeroth order Bessel function.  The first and second terms
in Eq.~(\ref{J}) are due to the Coulomb interaction $C$, where the
exchange term enters with a minus sign.  The parameter
$c=\sqrt{\pi/2}(e^2/\kappa a_{\rm B})/\hbar\omega_0$ ($\approx 2.4$, for
$\hbar\omega_0=3\,$meV) is the ratio between Coulomb and
confinement energy. The last term comes from the confinement potential
$W$.
Note that
typically $|J/\hbar\omega_0|\leq 0.2$. Also, we see
that $J>0$ for $B=0$, which must
be the case for a two-particle system that is time-reversal invariant.
The most remarkable feature of $J(B)$, however, is the change of sign
from positive to negative at $B=B_*^{\rm s}$, which occurs over a wide
range of parameters $c$ and $a$. This singlet-triplet crossing occurs
at about $B_*^{\rm s}=1.3\,{\rm T}$, for $\hbar\omega_0=3\,{\rm meV}$
($c=2.42$) and $d=0.7$. The transition from antiferromagnetic ($J>0$) to
ferromagnetic ($J<0$) spin-spin coupling with increasing magnetic
field is caused by the long-range Coulomb interaction, in particular
by the negative exchange term, the second term in Eq.~(\ref{J}).  As
$B\gg B_0$ ($\approx 3.5\,{\rm T}$ for $\hbar\omega_0=3\,$meV),
the magnetic field compresses the orbits by a factor $b\approx
B/B_0\gg 1$ and thereby reduces the overlap of the wavefunctions
exponentially. Similarly, the
overlap decays exponentially for $d\gg 1$.
Note however, that this exponential suppression is partly compensated
by the exponentially growing exchange term $\propto
\exp(2d^2(b-1/b))$. As a result, $J$ decays
exponentially as $\exp(-2d^2b)$ for large $b$ or $d$.
Thus,  $J$ can be tuned
through zero and then exponentially suppressed to zero by a magnetic field
in a
very efficient way (exponential switching is highly desirable to minimize
gate
errors). This sign reversal of
$J$ is due to the long-range  Coulomb forces
and is not contained in the standard Hubbard approximation which takes only
short-range interactions into account.
We  note that our Heitler-London approximation
breaks down explicitly (i.e. $J$ becomes negative even when $B=0$) for
certain
inter-dot distances when $c$ exceeds $2.8$.
By working around the magnetic field where {\em J} vanishes the
exchange interaction can be pulsed on, even without changing the tunneling
barrier
between the dots, either by an application of a local magnetic field, or by
exploiting a Stark electric field (which will also make the exchange
interaction
nonzero \cite{Burkard}).

Qualitatively similar results are obtained \cite{Burkard} when we refine
above Heitler-London result by taking into account higher levels and double
occupancy
of the dots (requiring a molecular orbit approach), and also for vertically
coupled
dots \cite{Seelig}.
Finally, we note that a spin coupling can also be achieved on a long
distance scale by using a cavity-QED scheme \cite{Imamoglu}.

\section{Quantum Communication with Electrons: Detection of
Entanglement \protect\cite{NoiseEntangle}}

The availability of pairwise entangled qubits---Einstein-Podolsky-Rosen
(EPR)
pairs \cite{Einstein}---is a necessary prerequisite in quantum
communication \cite{Bennett84}. The prime example of an EPR pair
considered here   is
the singlet state formed by two electron spins, its main feature being
its non-locality: If we separate the two
electrons   in real space, their total spin state
can still remain entangled. Such non-locality gives rise to
striking
phenomena such as violations of Bell inequalities and quantum
teleportation
and has been investigated for photons \cite{Aspect,Zeilinger},
but not yet for {\it massive} particles such as
electrons, let alone in a solid state environment. This is so because it is
difficult to first produce and to second detect entanglement of
electrons in a
controlled way.
In this section we  describe  an experimental set-up by which the
entanglement of electrons (once produced as described in the previous
section) can be
detected in noise measurements \cite{NoiseEntangle}, see Fig. 1.
The entangler is
assumed to be a device by which we
can generate (or detect) entangled electron states, a specific realization
being
above-mentioned double-dot system.  The presence of a beam
splitter ensures that the electrons leaving the entangler have a finite
amplitude to be interchanged (without mutual interaction). The quantity of
interest is then the current-current correlations (noise)
measured in  leads 3 and/or 4.
It is well-known \cite{Loudon} that particles with symmetric wave functions
show bunching behavior \cite{noise}
in the noise, whereas particles with antisymmetric wave functions show
antibunching  behavior. The latter situation is the one considered so
far for electrons in the normal state both in
theory \cite{Buettiker1,Martin} and in experiments \cite{Stanford,MartinSC}.
However, since the noise is produced by the charge degrees of freedom we
can expect \cite{DiVincenzoLossMMM} that in the absence of spin scattering
processes the noise is  sensitive to the symmetry of only the {\it
orbital part} of the wave function.   On the other hand, since  the spin
singlet of two electrons  is uniquely associated with a symmetric orbital
wave-function, and the three triplets with an antisymmetric
one we have thus a means to distinguish singlets from triplets through a
bunching or antibunching signature. Below we verify this expectation
explicitly, by
extending the standard scattering matrix approach \cite{Buettiker1,Martin} to
a
situation
with entanglement.  Finally, we also discuss the noise  which is produced by
spin
currents which can be present even in the absence of any charge currents.


The operator for the
current carried by electrons with spin $\sigma$ in lead $\alpha$ of a
multiterminal conductor can be written as \cite{Buettiker1,NoiseEntangle}
\begin{eqnarray}
  I_{\alpha\sigma}(t) =
\frac{e}{h\nu}\sum_{E,E'}&&\left[
a_{\alpha\sigma}^\dagger(E)a_{\alpha\sigma}(E')-b_{\alpha\sigma}^\dagger(E)
b_{\alpha\sigma}(E')
\right]
\exp\left[i(E-E')t/\hbar\right],\label{current_def}
\end{eqnarray}
where $a^\dagger_{\alpha \sigma} (E)$ creates an incoming electron in lead
$\alpha$
with spin $\sigma$ and energy $E$, and the operators $b_{\alpha\sigma}$ for
the
outgoing electrons
are related to the operators $a_{\alpha\sigma}$ for the incident electrons
via the
scattering matrix, $s_{\alpha\beta}$,
$ b_{\alpha\sigma}(E)=\sum_{\beta} s_{\alpha\beta}a_{\beta\sigma}(E)\, .$
We will assume that the scattering matrix is spin- and energy-independent.
Note that since we are dealing with discrete energy states here, we
normalize
the operators $a_{\alpha\sigma}(E)$
such that
$[a_{\alpha\sigma}(E),a_{\beta\sigma'}(E')^\dagger]=
\delta_{\sigma\sigma'}\delta_{\alpha\beta}\delta_{E,E'}/\nu$,
where  $\delta_{E,E'}$ is the Kronecker symbol, and $\nu$ the density of
states.
We assume that each lead consists of only a single quantum
channel.
We then obtain for the current
\begin{eqnarray}
  I_{\alpha\sigma}(t) = \frac{e}{h\nu}\sum_{E,E'} \sum_{\beta\gamma}
a_{\beta\sigma}^\dagger(E) A_{\beta\gamma}^\alpha
a_{\gamma\sigma}(E')e^{i(E-E')t/\hbar} ,\,\,\,\,
A_{\beta\gamma}^{\alpha} =
\delta_{\alpha\beta}\delta_{\alpha\gamma}
-s_{\alpha\beta}^{*} s_{\alpha\gamma}.\label{current}
\end{eqnarray}
We restrict ourselves here to unpolarized currents,
$I_\alpha=\sum_\sigma I_{\alpha\sigma}$.
The spectral density of the current fluctuations (noise)
$\delta I_{\alpha}=I_{\alpha}-\langle I_{\alpha}\rangle$
between the leads $\alpha$ and $\beta$ are defined as
\begin{equation}
  \label{cross1}
  S_{\alpha\beta}({\omega})
  = \lim_{T\rightarrow\infty}
  \frac{h\nu}{T}\int_0^T\!\!\!dt\,\,e^{i\omega t}
  \langle\Psi|\delta I_{\alpha}(t)\delta I_{\beta}(0)|\Psi\rangle,
\end{equation}
where the state $|\Psi\rangle$ is some arbitrary state to be specified
below.
Inserting the expression for the currents
Eq.~(\ref{current})
into this definition, we obtain for the zero frequency  correlations
\begin{eqnarray}
S_{\alpha\beta}
  = \frac{e^2}{h\nu} \sum_{\gamma\delta\epsilon\zeta}
    A_{\gamma\delta}^{\alpha}
A_{\epsilon\zeta}^{\beta}\sum_{E,E',\sigma\sigma'}
    &&\left[\langle\Psi| a_{\gamma\sigma}^\dagger(E) a_{\delta\sigma}(E)
    a_{\epsilon\sigma'}^\dagger(E') a_{\zeta\sigma'}(E')
|\Psi\rangle \right. \nonumber \\
&&\left.    - \langle\Psi| a_{\gamma\sigma}^\dagger(E)
a_{\delta\sigma}(E)
|\Psi\rangle
\langle\Psi| a_{\epsilon\sigma'}^\dagger(E') a_{\zeta\sigma'}
(E') |\Psi\rangle \right].
\label{cross3}
\end{eqnarray}
We note that since $|\Psi\rangle$ in general does not describe a Fermi
liquid
state, it is
not possible to apply Wick's theorem.

We will now investigate the noise correlations for scattering with the
entangled incident state $|\Psi\rangle=|\pm \rangle$, where
$  |\pm\rangle
  = \frac{1}{\sqrt{2}}\left( a_{2\downarrow}^\dagger(\epsilon_2)
a_{1\uparrow}^\dagger(\epsilon_1)
    \pm a_{2\uparrow}^\dagger(\epsilon_2)
a_{1\downarrow}^\dagger(\epsilon_1)\right) |0\rangle$ .
The state $|-\rangle$ is the spin singlet, $|S\rangle$,
while $|+\rangle$ denotes one of the spin triplets
$|T_{0,\pm}\rangle$; in the following we will present a calculation of
the noise for $|+\rangle=|T_0\rangle$, i.e. the triplet with $m_z=0$.

Substituting $|\pm\rangle$
for $|\Psi\rangle$,
we get
$ \langle\pm|\delta I_{\alpha} \delta I_{\beta}|\pm\rangle
  =  \langle\uparrow\downarrow|\delta I_{\alpha} \delta I_{\beta}|
  \uparrow\downarrow\rangle
  \pm\langle\uparrow\downarrow|\delta I_{\alpha} \delta I_{\beta}|
  \downarrow\uparrow\rangle$,
where the upper (lower) sign of the exchange term refers to triplet
(singlet).
After some straightforward manipulations,
we obtain the following result for the
correlations between the leads $\alpha$ and $\beta$,
  $S_{\alpha\beta}
   = \frac{e^2}{h\nu}\left[\sum_{\gamma\delta}\!{}^{'}
    A_{\gamma\delta}^{\alpha}A_{\delta\gamma}^{\beta}
   \mp \delta_{\epsilon_1,\epsilon_2}
    \left(A_{12}^{\alpha}A_{21}^{\beta} +A_{21}^{\alpha}A_{12}^{\beta})
\right)\right]$,
where $\sum_{\gamma\delta}^{\prime}$ denotes the sum over $\gamma=1,2$ and
all $\delta\neq\gamma$, and where again the upper (lower) sign refers to
triplets (singlets).

\begin{figure}
  \parbox{4cm}{\psfig{file=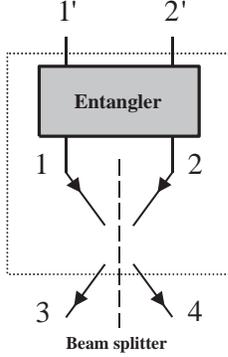,width=3cm}}
  \hfill
  \parbox{11cm}{\caption{
Uncorrelated electrons are fed into the entangler $E$ through the Fermi
leads $1'$ and $2'$. The entangler
is a device (see text) that produces pairs of electrons in the entangled
spin singlet or one of the spin triplets
and injects one of the electrons into lead $1$ and the other into
lead $2$.  For singlets
we get bunching due to their orbital symmetry, whereas for triplets
we get antibunching due to their orbital antisymmetry.
Note
that instead of being divided into an entangler and a beam splitter, the
whole area
inside the dotted box can be viewed as a two-particle scattering
region.\hfill
\label{fig1}}}
\end{figure}

We apply above formula
now to the set-up
shown in Fig.~\ref{fig1}
involving four leads, described by the  scattering matrix elements,
$s_{31}=s_{42}=r$, and $s_{41}=s_{32}=t$,
where $r$ and $t$ denote the reflection and transmission amplitudes at the
beam splitter, resp., and with no
backscattering,
$s_{12}=s_{34}=s_{\alpha\alpha}=0$.
The unitarity of the s-matrix implies $|r|^2+|t|^2=1$, and Re$[r^*t]=0$.
Using above relations, we obtain finally \cite{footnote1},
\begin{equation}
  \label{noise}
  S_{33}=S_{44}=-S_{34}=2\frac{e^2}{h\nu}T\left(1-T\right)
  \left(1\mp \delta_{\epsilon_1,\epsilon_2}\right),
\end{equation}
where $T=|t|^2$ is the probability for transmission through the
beam splitter.
The calculation for the remaining two triplet states
$|+\rangle=|T_\pm\rangle=|\uparrow\uparrow\rangle
,|\downarrow\downarrow\rangle$
yields the same result Eq.~(\ref{noise}) (upper sign).
Note that the total current $\delta I_3 + \delta I_4$ does note fluctuate,
{\it i.e.}
$S_{33}+S_{44} +2S_{34}=0$, since we have excluded backscattering.
For the average current in lead $\alpha$ we obtain
$\left|\langle I_\alpha\rangle\right| = e/h\nu$,
with no difference
between singlets and triplets.
Then, the Fano factor
$F = S_{\alpha\alpha} /\left|\langle I_\alpha\rangle\right|$
takes the following form
\begin{equation}
  \label{fano}
  F =  2eT(1-T)\left(1\mp \delta_{\epsilon_1,\epsilon_2}\right),
\end{equation}
and correspondingly for the cross correlations.
This  result confirms our expectation stated in the introduction: if two
electrons
with the same energies, $\epsilon_1=\epsilon_2$, in the singlet
state $|S\rangle = |-\rangle$ are injected into the leads $1$ and $2$,
then the zero frequency noise is {\it enhanced} by a factor of two,
$F=4eT(1-T)$, compared to the shot noise of uncorrelated particles,
$F=2eT(1-T)$. This enhancement of noise is
due to {\it bunching} of electrons in the outgoing leads, caused by the
symmetric orbital wavefunction of the spin singlet $|S\rangle$.
On the other hand, the triplet states $|+\rangle = |T_{0,\pm}\rangle$
exhibit an {\it antibunching} effect, leading to a complete
suppression of the zero-frequency noise in Eq.~(\ref{noise}),
$S_{\alpha\alpha}=0$.
The noise enhancement for the singlet $|S\rangle$ is a
unique signature for entanglement (there exists no unentangled state with
the same symmetry), therefore entanglement can be observed by
measuring the noise power of a mesoscopic conductor.
The triplets can be further distinguished from each other
if we can measure the
spin of the two electrons in the outgoing leads, or if we insert
spin-selective tunneling devices \cite{Prinz} into leads 3,4
which would filter a certain spin polarization.

We emphasize that above results remain unchanged if we consider
states $|\pm \rangle$ which are created above a Fermi sea. We have shown
elsewhere \cite{DiVincenzoLossMMM} that the entanglement of two electrons
propagating
in a Fermi sea gets reduced by the quasiparticle weight $z_F$ (for each
lead one factor)
due to the presence of interacting electrons.
In the metallic regime $z_F$ assumes typically some finite
value \cite{qpweight},
and thus as long as spin scattering processes are small the above
description
for non-interacting electrons remains
valid.

Finally we discuss the current noise induced by the spin transport
in a two-terminal conductor attached to Fermi leads with spin-dependent
chemical potentials $\mu_{\sigma}$. From Eq.(\ref{current})
we immediately obtain $\left<I_{\sigma}\right>={e\over h}T\Delta
\mu_{\sigma},$
where we have introduced the difference of chemical potentials
$\Delta\mu_{\sigma}=\mu_{1\sigma}-\mu_{2\sigma}$ for each
spin orientation $\sigma$. Then, applying
Wick's theorem to Eq.\ (\ref{cross3}) we obtain the noise power
$S={{e^2}\over h}T(1-T)\left(\left|\Delta\mu_{\uparrow}\right
|+\left|\Delta\mu_{\downarrow}\right|\right).$
Thus, the contribution of two spin subsystems to the noise
is independent, as it should be if there is no spin interaction.
Therefore, we can rewrite above expression as
$S=e(1-T)\left(\left|\left<I_{\uparrow}\right>\right|+\left|\left
<I_{\downarrow}\right>\right|\right)$.
In particular, when $\Delta\mu_{\uparrow}=\Delta\mu_{\downarrow}$
we obtain the usual result \cite{noisesuppression} $S=e(1-T)\left|I_c\right|$
for the shot noise induced by the charge current
$I_c\equiv\left<I_{\uparrow}\right>+\left<I_{\downarrow}\right>$.
On the other hand, we can consider the situation where
$\Delta\mu_{\uparrow}=-\Delta\mu_{\downarrow}$, and thus there
is no charge current through the
conductor, $I_c=0$. Still, there is a non-vanishing spin current
$I_s\equiv\left<I_{\uparrow}\right>-\left<I_{\downarrow}\right>$, and
according
to the above result one can
observe the current noise $S=e(1-T)\left|I_s\right|$
induced by spin transport only.

We finally mention a further scenario where entanglement can be measured
in the current \cite{LossSukhorukov}: A double-dot system which is weakly
coupled to an ingoing (1) and an outgoing (2) lead held at chemical
potentials
$\mu_{1(2)}$, but where now an electron coming from lead 1 (2) has the
option
to tunnel into {\it both} dots 1 and 2 with amplitude $\Gamma$. This results
in
a closed loop, and applying a magnetic field, an Aharonov-Bohm phase
$\varphi$
will be picked up by an electron traversing the double-dot. In the Coulomb
blockade regime we find that due to cotunneling the current traversing the
double-dot becomes  (for $U>|\mu_1\pm \mu_2|>J>k_BT, 2\pi \nu \Gamma^2$,
with U
the single-dot charging energy)
\begin{equation}
I=e\pi\nu^2\Gamma^4{{\mu_1-\mu_2}\over {\mu_1\mu_2}}(2 \pm\cos
\varphi),
\end{equation}
where the upper (lower) sign refers
to triplet (singlet) state in the double-dot. Similarly,
we find for the noise, $S_\omega\propto (2
\pm\cos \varphi)$.
Thus, the current and noise reveal whether the double-dot  is in a singlet
or
triplet state.
The triplets can be further distinguished by applying spatially
inhomogeneous
magnetic fields leading to a beating of the AB phase oscillation due to the
Berry phase \cite{LossSukhorukov}.

\begin{thebibliography}{99}
\bibitem{footnote1}
For finite frequencies, we obtain the noise power
$S_{\alpha\alpha}(\omega) = (2e^2/h\nu)[(1-\delta_{\omega,0})
+2T(1-T)(\delta_{\omega,0}\mp\delta_{\omega,(\epsilon_1-\epsilon_2)/\hbar})]
$.
\bibitem{qpweight}
For instance, in metals such as bulk Cu the quasiparticle weight becomes
within
RPA approximation
$z_F=0.77$, while for a GaAs 2DEG we find\protect \cite{DiVincenzoLossMMM}
(also
within RPA) $z_F=1-r_s(1/2+1/\pi)=0.66$ for the GaAs interaction parameter
$r_s=0.61$.
\end{thebibliography}
\end{document}